\documentclass[letterpaper,twocolumn,10pt]{article}

\usepackage{zhanggroup}
\usepackage[absolute]{textpos}
\usepackage{amsmath}
\usepackage{tikz,eso-pic}
\usepackage{amsfonts}
\usepackage{xspace} 
\usepackage{mathrsfs}
\usepackage{amsmath}
\usepackage{amsthm}
\usepackage{subcaption}
\usepackage{booktabs}
\usepackage{multirow}
\usepackage{graphicx}
\usepackage{flushend}
\usepackage{url}
\usepackage{xurl}
\usepackage{caption, subcaption}
\usepackage{float}
\usepackage[normalem]{ulem}
\usepackage{hyperref}
\hypersetup{
  colorlinks,
  linkcolor={blue!70!green},
  citecolor={green!70!blue},
  urlcolor={orange!70!red}
}
\usepackage[encapsulated]{CJK}
\newcommand{\chinese}[1]{\begin{CJK}{UTF8}{gbsn}#1\end{CJK}}

\newcommand{\system}{Mondrian\xspace}
\newcommand{\name}{prompt abstraction\xspace}
\newcommand{\mypara}[1]{\smallskip\noindent{\bf {#1}.}\xspace}

\begin{document}

\date{}

\title{\Large \bf Mondrian: Prompt Abstraction Attack Against Large Language Models for Cheaper API Pricing}

\author{
{\rm Wai Man Si},\ \ \
{\rm Michael Backes},\ \ \
{\rm Yang Zhang}
\\
\\
\textit{CISPA Helmholtz Center for Information Security}\ \ \
}

\maketitle

\begin{abstract}

The Machine Learning as a Service (MLaaS) market is rapidly expanding and becoming more mature.
For example, OpenAI's ChatGPT is an advanced large language model (LLM) that generates responses for various queries with associated fees.
Although these models can deliver satisfactory performance, they are far from perfect. 
Researchers have long studied the vulnerabilities and limitations of LLMs, such as adversarial attacks and model toxicity.
Inevitably, commercial ML models are also not exempt from such issues, which can be problematic as MLaaS continues to grow.
In this paper, we discover a new attack strategy against LLM APIs, namely the \name attack.
Specifically, we propose \system, a simple and straightforward method that abstracts sentences, which can lower the cost of using LLM APIs.
In this approach, the adversary first creates a pseudo API (with a lower established price) to serve as the proxy of the target API (with a higher established price).
Next, the pseudo API leverages \system to modify the user query, obtain the abstracted response from the target API, and forward it back to the end user.
Our results show that \system successfully reduces user queries' token length ranging from 13\% to 23\% across various tasks, including text classification, generation, and question answering.
Meanwhile, these abstracted queries do not significantly affect the utility of task-specific and general language models like ChatGPT.
\system also reduces instruction prompts' token length by at least 11\% without compromising output quality.
As a result, the \name attack enables the adversary to profit without bearing the cost of API development and deployment.

\end{abstract}

\section{Introduction}
\label{sec:introduction}

Natural Language Processing (NLP) has been a popular research topic for a long time because it can make human life more convenient, such as assisting in writing emails and summarizing news articles. 
Despite its popularity, real-world NLP applications have been relatively scarce due to their limited capabilities.
However, recent advancements in deep learning have changed this situation. 
For instance, ChatGPT is a conversational language model product capable of solving diverse problems, from writing emails and generating code to passing the bar exam~\cite{SBHP23}. 
Developing such products requires a substantial amount of effort and resources.
They need a huge amount of high-quality data for training, which involves costly human labeling, as well as extensive computational resources for deployment.
Not to mention the engineering effort behind the product, such as hardware and network infrastructure.
Therefore, it is common for users to use existing services instead of building their own. 
Companies also provide fine-tuning services, allowing users to create their own custom models. 
We can expect an increasing number of commercial NLP services as it continues to grow.

Since providing such services is expensive, companies would charge users based on their usage to balance the cost and generate revenue.
Typically, the pricing mechanism of using LLM APIs is based on factors such as the token or character length of the input and output.
For example, ChatGPT is priced \$0.002 per 1,000 tokens,\footnote{\url{https://openai.com/pricing}} and Google's ``text-bison'' costs \$0.001 per 1,000 characters.\footnote{\url{https://cloud.google.com/vertex-ai/pricing}}
It can cost a small business over \$21,000 a month using ChatGPT~\cite{CZZ23}.
Consequently, the pricing mechanism and the cost of using LLM APIs open up an opportunity for maliciously profiting.
In this paper, we introduce a novel attack strategy named the \name attack. 
This is the first attack specifically designed to target the pricing mechanisms of NLP APIs.

\mypara{Attack Methodology}
To perform the \name attack, the adversary first offers a pseudo API to the public with a lower established price than the target API, which can attract users because of the price discrepancy.
Once the pseudo API receives the query from the user, \system abstracts the query to a certain extent before forwarding it to the target API to obtain the corresponding response at an even lower cost.
\system features two operations: Delete and Transform.
The former removes words that are not necessary, while the latter converts words into shorter equivalents in terms of token or character length. 
Thus, \system can abstract the user query without significantly altering the semantic, ensuring that the response from the target API is not largely impacted.
This allows the adversary to profit from the pricing difference between the pseudo API and the target API.
Given millions of users and queries a day, even a 1\% profit can be significant.

Besides, it is important to highlight the potential risks for users and companies under a successful attack.
This attack can cause privacy risks for the user since the adversary has access to the query.
For example, the adversary can capture sensitive information if the user sends such data to the pseudo API.
In April 2023, Samsung employees accidentally leaked trade secrets via ChatGPT.\footnote{\url{https://www.bloomberg.com/news/articles/2023-05-02/samsung-bans-chatgpt-and-other-generative-ai-use-by-staff-after-leak}}
Moreover, since the adversary also has access to the response, they can modify or replace it with unsafe content.
The adversary can also inject advertisements into the response for potential profit.
Another risk involves parasitic computing, wherein the adversary can offer the exact API without incurring the cost of building and hosting the API.
For instance, Training GPT-3 costs \$4.6 million\footnote{\url{https://lambdalabs.com/blog/demystifying-gpt-3}} and running it costs around \$700,000 a day.\footnote{\url{https://www.semianalysis.com/p/the-inference-cost-of-search-disruption}}

\mypara{Evaluation}
To demonstrate the efficacy of the \name attack, we conduct experiments on different model setups, including task-specific and general language models.
For task-specific models, given that users may upload or fine-tune their models for specific tasks using online platforms such as Google Cloud or OpenAI, we evaluate our attack in a simulated environment using public models from Huggingface Hub~\cite{WDSCDMCRLFB19}.
Our results indicate that abstracted sentences generated by \system deliver strong performance across various NLP tasks.
Specifically, it has less than a 1\% performance drop in text classification while reducing the token length of sentences ranging from 11\% to 23\%.
For text generation, we perform experiments on the text summarization task using CNN/DailyMail and XSum.
The content of generated summaries based on original and abstracted inputs are similar, with a 15\% reduction in the number of input tokens.
Furthermore, our attack achieves a 15\% token reduction for the question-answering (QA) task with only a negligible drop in performance.
For general language models, we evaluate our attack against the recently popular product, ChatGPT. 
Similarly, we use the same abstracted data from \system and achieve strong performance. 
Furthermore, we employ Awesome ChatGPT Prompt (ACP) and Alpaca datasets to simulate real-world user input and maintain comparable performance while reducing the number of tokens by 11\%.

\section{Preliminaries}
\label{sec:background}

To evaluate our \name attack, we use text classification, generation, and QA tasks.
To provide a comprehensive understanding, we briefly introduce each of these tasks.
Then, we introduce MLasS along with its pricing mechanism.

\subsection{Text Classification}

Text classification is widely recognized and extensively employed in real-world applications. 
For example, Perspective API\footnote{\url{https://perspectiveapi.com/}} and OpenAI's Moderation API are two APIs that provide attribute analysis, such as toxicity detection. 
Generally, when presented with the input sentence, the model produces the corresponding output in either a continuous number or a discrete label, i.e., sentiment or topics.
Moreover, the model can also handle tasks such as sentence matching and inference by accepting two input sentences. 
The model is expected to understand the relationship between two input sentences in such cases. 
Recently, text classification can be formulated as generation using general language models, often in zero-shot or few-shot settings~\cite{BMRSKDNSSAAHKHCRZWWHCSLGCCBMRSA20, MLHALHZ22}.

\subsection{Text Generation}

Text generation is one of the most prevalent NLP tasks in real-world applications. 
This task can be broadly classified into two types: guided and unguided text generation. 
In guided text generation, the model is required to produce output that is closely related to the input.
A common example of guided text generation is summarization, which automatically generates summaries for news articles or other text content.
It requires the model to generate output that is highly coherent with the input.
On the other hand, unguided text generation allows for greater flexibility in terms of output context and format. 
For instance, it is common to prompt the model to generate recommendations, search for specific information, and entertain the user.
It is often that the answer can be open-ended.

\subsection{Question Answering}

Question answering (QA) is an NLP task that involves finding answers based on the given question and contextual information. 
Typically, QA is solved using the extractive setup, where the model is expected to extract the answer from the context in response to the question. 
It requires the model to predict the start and end location within the given context.
Some datasets like SQuAD 2.0 intentionally include unanswerable questions within the context to assess the model's capabilities~\cite{RJL18}.
Furthermore, QA can also be performed in the generative setup, which is similar to text generation by querying the model with the context and question.
Similarly, advanced general language models have demonstrated the ability to generate answers even without explicit context with their own knowledge.

\subsection{Commmerical NLP}

Machine learning as a service (MLaaS) includes diverse services, including model training and deployment. 
Due to the massive computational resources and domain expertise required to train specific ML models, individuals or small businesses often choose intermediary or existing services instead of building their own from scratch.
For example, OpenAI allows users to fine-tune OpenAI's pre-trained models for their own purposes, with costs ranging from \$0.0016 to \$0.12 per 1,000 tokens, depending on the user's requirements.
Besides, both OpenAI and Google provide general language models that can be directly used to solve different problems.
ChatGPT is priced \$0.002 per 1,000 tokens, and Google's ``text-bison'' costs \$0.001 per 1,000 characters.
It is clear that the length of inputs determines the cost of using commercial APIs.
In essence, longer sentences result in higher costs, and this cost structure creates the attack potential for our proposed \name attack with \system.

\section{Threat Model and Methodology}
\label{sec:methodology}

\begin{figure*}[!t]
\centering
\includegraphics[width=1.7\columnwidth]{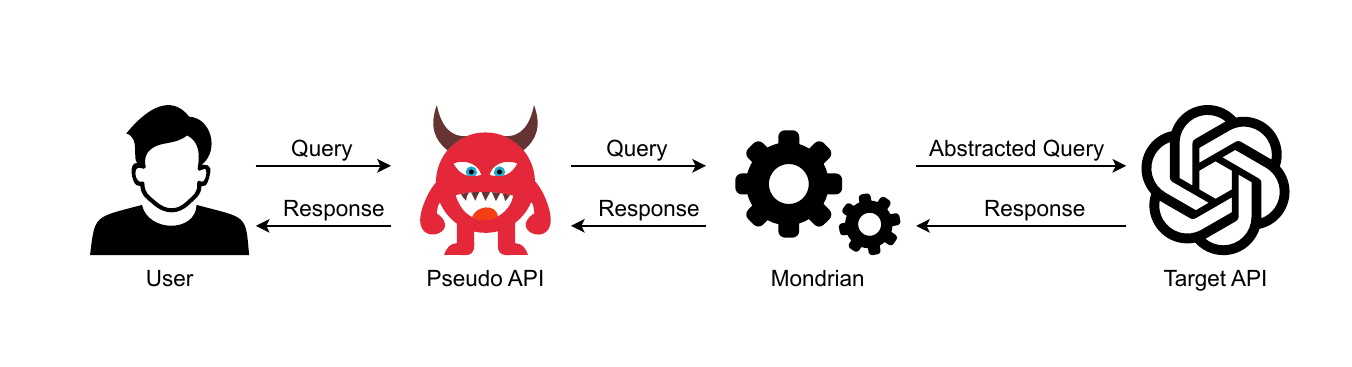}
\caption{An overview of the \name attack pipeline.}
\label{fig:pipeline}
\end{figure*}

\subsection{Threat Model}

\mypara{Adversary's Goal}
The \name attack is a new type of attack strategy that occurs during the testing time. 
The primary goal of an adversary is to generate profits by offering the pseudo API to the public. 
The adversary constructs the pseudo API as a proxy and leverages the target API instead of developing its own API.
Ideally, the pseudo API obtains the proper response from the target API, with the modified (abstracted) user query through \system.
This approach enables the pseudo API to query the target API at a lower cost than using the original user query. 
Therefore, the adversary can offer the pseudo API at a higher price while still keeping it below the established price of the target API.
We focus on tack-specific and general language models given their practical use cases and recent popularity. 
Through a successful attack, the adversary can profit from the price difference while saving the costs of building and maintaining the API.

\mypara{Adversary's Background Knowledge}
We assume that the adversary has direct access to the target API. 
However, the target API may have different restrictions, including a safety filter and rate limits, which would also apply to the pseudo API.
To fulfill the adversary's objective, \system needs to modify the user query in terms of the sentence's token length.
For this purpose, the adversary needs a tokenizer, which breaks the sentence into individual tokens, such as words or subwords, allowing them to measure the sentence's token length.
The adversary also has access to a function capable of performing sentence similarity measures and uses WordNet for searching for synonyms~\cite{M95}.
Importantly, \system does not require any training, meaning the adversary does not need any training or testing data. 
Furthermore, it is only aware of the user's query once received.

\subsection{Attack Methodology}

In the \name attack, the abstraction of the user query is the key for the attack to succeed.
In this section, we introduce our \system to obtain the abstracted user query.
Essentially, \system applies a series of operations that iteratively manipulate the user query, resulting in the abstracted version of the query.
We begin by providing an overview of the \name attack pipeline. 
Then, we detail each component of \system, including tokenization and each operation. 
Finally, we present a summary of \system abstraction process.

\mypara{Attack Pipeline}
To launch the \name attack, the adversary sets up the pseudo API with \system, denoted as $f'$.
Next, the adversary selects the target API $f$, which can be either a task-specific or general language model.
The pseudo API functions act as a pure proxy and place \system between the pseudo and target API.
When the pseudo API receives the user query, denoted as $x = x_1 x_2 ... x_n$, \system modifies the original query $x$ into the abstracted query $x'$.
This abstraction process has two crucial requirements.

The first requirement is that \system should minimize the length of the original sentence in terms of its tokens or characters.
Consequently, the length of the abstracted sentence is lesser than the original sentence as $|T(x)| > |T(x')|$, where $T$ is the tokenizer that is similar or the same to the target API's tokenizer.
Ideally, querying with the abstracted sentence will cost lower than the original sentence on the target API.

The second requirement is that the abstracted sentence should maintain semantic coherence with the original sentence.
However, the grammar or structure does not need to be correct, as we have discovered that LLMs (target APIs) are robust to understand it.
To ensure that the central meaning of the abstracted sentence does not significantly deviate from the original sentence, we measure the sentence similarity between them via function $D$.
For the function, we can use different methods, such as cosine similarity or Jaccard similarity.
If the semantic deviates too much, the target API may fail to generate the correct output. 
Ideally, responses from the target API should exhibit high similarity between using the original query and the abstracted query: $f(x) \approx f'(x')$.

In addition, we do not use the target API to assist us in the abstraction to minimize the costs associated with accessing the target API.
Once the abstraction process is complete, the pseudo API forwards the abstracted query to the target API.
The final stage of the process involves the pseudo API simply replying to the user with the response received from the target API.

\mypara{Tokenization}
In commercial NLP services, pricing is usually determined by the length of the sentence, which is measured in tokens or characters. 
While the measurement of characters is straightforward, tokens are obtained through tokenization.
Tokenization divides the raw text into smaller segments called tokens using a tokenizer. 
These tokens can be subwords or complete words and play a vital role in helping NLP models interpret the meaning of the text.
A good tokenization can significantly enhance the performance of the model.

Traditionally, tokenization relies on splitting the text based on whitespace and punctuation. 
However, many state-of-the-art models have adopted subword tokenization in recent years.
This approach breaks words into subword units, enabling a more robust text representation. 
In subword tokenization, frequently occurring words are retained entirely, while less common words are split into subwords. 
A common example is that the word ``few'' is kept due to its high frequency, while rarer words like ``fewer'' and ``fewest'' are split into subwords such as ``few,'' ``er,'' and ``est.''
This approach prevents the language model from identifying ``fewer'' and ``fewest'' as different words.
Moreover, subword tokenization can reduce the chance of treating words as unknown during training.
Importantly, it also provides a natural environment for our attack, as some words can be replaced by their synonyms with fewer tokens without changing the meaning.

\mypara{Operations}
To achieve the abstracted sentence, we employ two simple operations: Delete and Transform. 
The Delete operation involves the removal of individual words (and punctuations) from the sentence.
As shown in \autoref{sec:ablation}, this operation proves effective in abstracting the sentence.
It allows us to eliminate redundant words without harming the target model to understand the query and generate the correct response.
For example, the query ``How do I make an HTTP request in Javascript?'' can be abstracted to ``HTTP requests in Javascript'' by eliminating unnecessary words while preserving the core idea of the query.

Next, we have the Transform operation, which transforms specific words into their synonyms that come with fewer tokens, utilizing WordNet.
The essential advantage of this operation is that it retains the original meaning of the sentence, ensuring no compromise in semantics.
For instance, the query ``HTTP requests in Javascript'' maintains semantic equivalence with ``HTTP request in Javascript'' even though the grammar is slightly incorrect.
By employing the latter, we can reduce the cost of using the target API.
Further examples of the Transform operation include substitutions such as using the abbreviation ``US'' instead of the full form, ``United States.''
Such transformation can reduce the sentence's character length significantly, resulting in a huge reduction in the cost.

\mypara{Abstraction Process}
Overall, when any user queries the pseudo API, \system begins by modifying the query $x$ into its abstract version $x'$.
Initially, \system splits the query into sentences if it is a paragraph.
In this work, our \system does not consider the relation between sentences since it should not remove essential words but just irrelevant words.
We leave further study on it in the future.
Next, each word in the sentence is systematically processed by the Delete and Transform operations during each iteration.
Throughout each iteration, \system generates an array of candidates.
It then calculates an objective score for each prospective candidate $x'$ as follows:
\begin{align*}
    O_{length} = |T(x')|
\end{align*} 
In this paper, we adopt a greedy search strategy, selecting the candidate with the lowest objective as the best candidate to proceed into the next iteration.
However, we must ensure that the semantics of the prospective sentence remain consistent with the original sentence.
We measure the semantic distance by computing their cosine similarity within the embedding space. 
To maintain the semantics, we pre-define a threshold $\alpha$, which is a hyperparameter. 
The abstraction process stops if the cosine similarity between the original sentence and the candidate exceeds the threshold: $D(x, x') > \alpha$.
After abstraction, our pseudo API obtains the response by querying the target API with the best candidate- the abstracted sentence- and then forwards it to the user.
Importantly, our abstraction is highly efficient as it only takes $O(|x|)$ to run at the worst case.
Furthermore, only the similarity measure component requires GPU computations, which are significantly less resource-intensive than operating the target API, i.e., hosting a large-scale language model.

\mypara{Comparison with Existing Cost Reducing Methods}
Recent studies have introduced several methods aimed at reducing the cost associated with API usage~\cite{CZZ20, CZZ23}.
Chen et al.~\cite{CZZ23} propose using a learning-based technique to dynamically assign different queries to combinations of APIs for each dataset and task.
Their method has demonstrated promising results in terms of cost reduction and task performance. 
However, their method requires access to the dataset and training a model with the dataset.
From an attacker's perspective, acquiring prior access to the user's query is intractable. 
In contrast, our method does not require access to the user's query, making our attack approach more applicable in real-world scenarios.
Nevertheless, it is worth noting that our pseudo API can blend into their API combinations seamlessly to enhance performance further.

\section{Experimental Settings}
\label{sec:settings}

This section provides an overview of the experimental setup used to evaluate \system. 
First, we introduce the evaluation dataset, followed by the implementation of the target model and pseudo API.
Then, we explain the method and metrics used to evaluate \system.

\subsection{Evaluation Dataset}

\mypara{Text Classification}
We use single-sentence and inference tasks to study the effectiveness of \system.
First, we introduce datasets that include a single input sentence below:

\begin{itemize}
\item \textbf{SST-2}~\cite{WSMHLB19} contains sentences extracted from movie reviews with human annotations indicating their sentiment as either positive or negative.
\item \textbf{AGnews}~\cite{ZZL15} includes news articles covering a wide range of topics such as world events, sports, business, and science \& technology.
\item \textbf{IMDB}~\cite{MDPHNP11} consists of lengthy movie reviews and the sentiment label in positive or negative.
\end{itemize}

Second, we introduce datasets that come with two input sentences.
It requires the model to understand the relationship between sentences, which is more challenging for the model to understand between two abstracted input sentences.

\begin{itemize}
\item \textbf{QNLI}~\cite{WSMHLB19} comprises question-answering pairs, challenging the task of determining whether the context sentence aligns with the provided answer.
\item \textbf{MNLI}~\cite{WSMHLB19} contains crowdsourced sentence pairs with textual entailment annotations.
The task involves predicting the relationship between a premise sentence and a hypothesis sentence, whether it is entailment, contradiction, or neutral.
\item \textbf{RTE}~\cite{WSMHLB19} originate from annual competitions focused on textual entailment. 
They are transformed into a two-class classification task, distinguishing between entailment and not entailment.
\end{itemize}

\mypara{Text Generation}
For text generation, we utilize summarization as a testing bed.
Summarization involves the task of condensing long inputs, such as articles, into shorter versions. 
It allows us to assess the effectiveness of \system in handling long paragraphs and examine if the target model still works with abstracted inputs.

\begin{itemize}
\item \textbf{CNN/DailyMail (CNNDM)}~\cite{HKGEKSB15} comprises news articles written by journalists from CNN and the Daily Mail.
\item \textbf{XSum}~\cite{NCL18} is designed for generating a single sentence summary. 
It contains news articles from the BBC articles.
\end{itemize}

\mypara{Question Answering}
The question-answering task requires the model to comprehend the question and extract the answer from the given context. 
We assess \system's ability to abstract both the context and the question while preserving essential information.

\begin{itemize}
\item \textbf{SQuAD2}~\cite{RJL18} is a combination of questions sourced from SQuAD1 and unanswerable questions crafted by crowd workers~\cite{RZLL16}.
\end{itemize}

\mypara{Instrcution}
Previous tasks often come with a specific structure for the input and output, suitable for task-specific models. 
However, it is common for users to query the model with general requests such as writing emails or role-playing scenarios such as acting as a Plagiarism Checker. 
To evaluate our \system's performance in real-world usage, we employ two different instruction datasets to simulate actual use cases.

\begin{itemize}
\item \textbf{Awesome ChatGPT Prompts (ACP)} is a collection of prompt examples to be used with the ChatGPT model released on GitHub.\footnote{\url{https://github.com/f/awesome-chatgpt-prompts}}
\item \textbf{Alpaca}~\cite{alpaca} comprises instructions and demonstrations generated by OpenAI's ``text-davinci-003'' engine, and its primary purpose is to facilitate instruction-tuning for language models originally.
\end{itemize}

Due to the expenses of using ChatGPT, we randomly selected 500 samples for classification and 100 samples each for summarization and QA for the experiment.
As for the instruction dataset, we sampled 500 data since there is no definitive ground truth, and a larger data pool allows for a more comprehensive evaluation and analysis.
In addition, we also summarize the statistics for all of these datasets in~\autoref{app:dataset}.

\subsection{Target Models}

In this work, we employ two distinct types of models: task-specific and general language models.
For the task-specific model, we consider the target API to be a fine-tuned model or a model hosted on a commercial platform. 
To create our testing environment, we utilize various public models available on HuggingFace Hub (full details provided in~\autoref{app:target}).
The purpose of this setup is to simulate a scenario where the user or company uploads their private models or fine-tunes pre-trained models for particular tasks.
For instance, Google Cloud offers a range of models for analyzing text, such as entity and sentiment recognition.
For the general language model, we use ChatGPT, which is a popular commercial AI product developed by OpenAI.
Unlike task-specific models, the general language model is capable of addressing a wide range of tasks without fine-tuning.
Our study covers most existing NLP models in research and commercial environments with both model settings.

\begin{figure*}[!t]
\centering
\begin{subfigure}[!t]{0.6\columnwidth}
\centering
\includegraphics[width=\columnwidth]{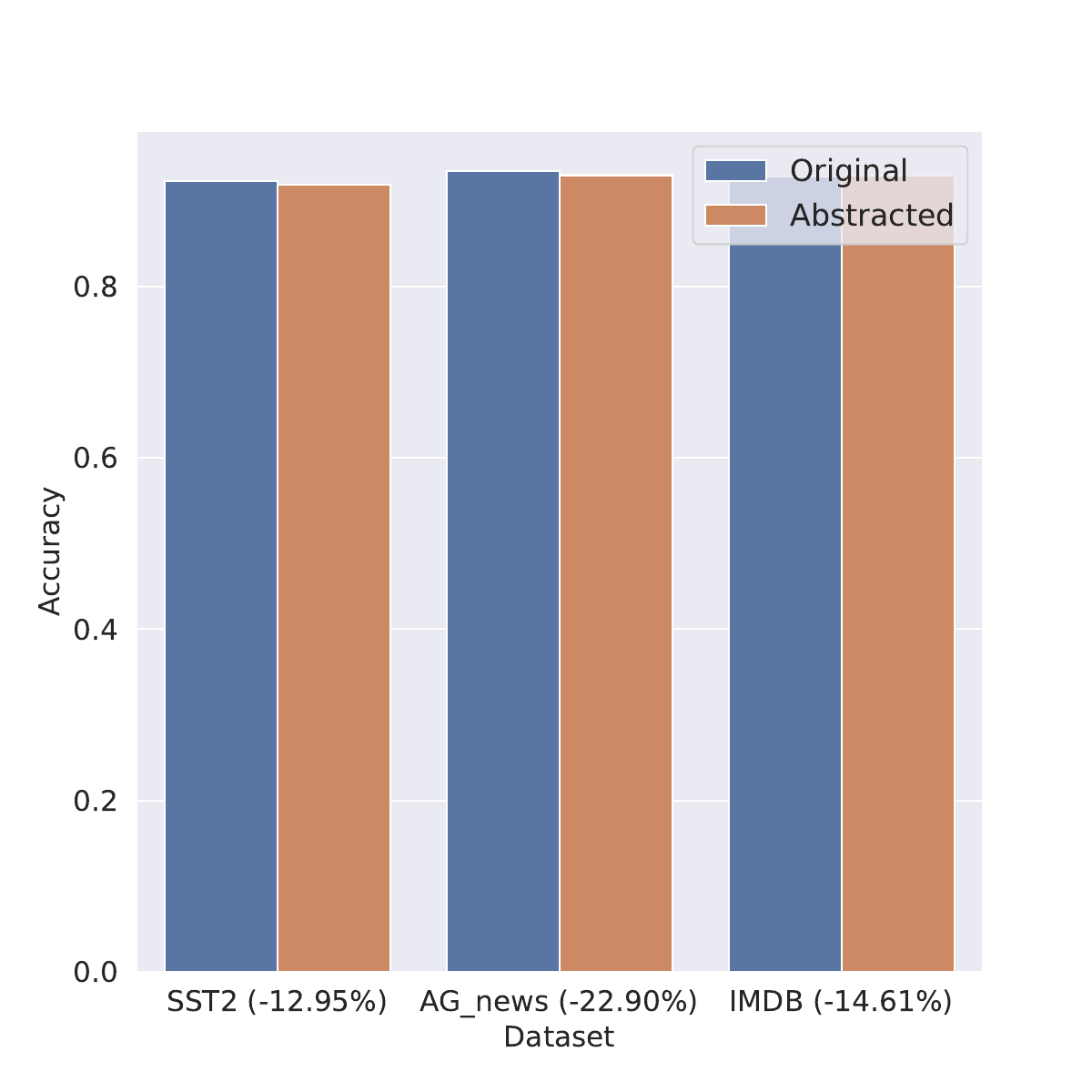}
\caption{Single-Sentence Tasks}
\label{fig:cls_exp}
\end{subfigure}
\begin{subfigure}[!t]{0.6\columnwidth}
\centering
\includegraphics[width=\columnwidth]{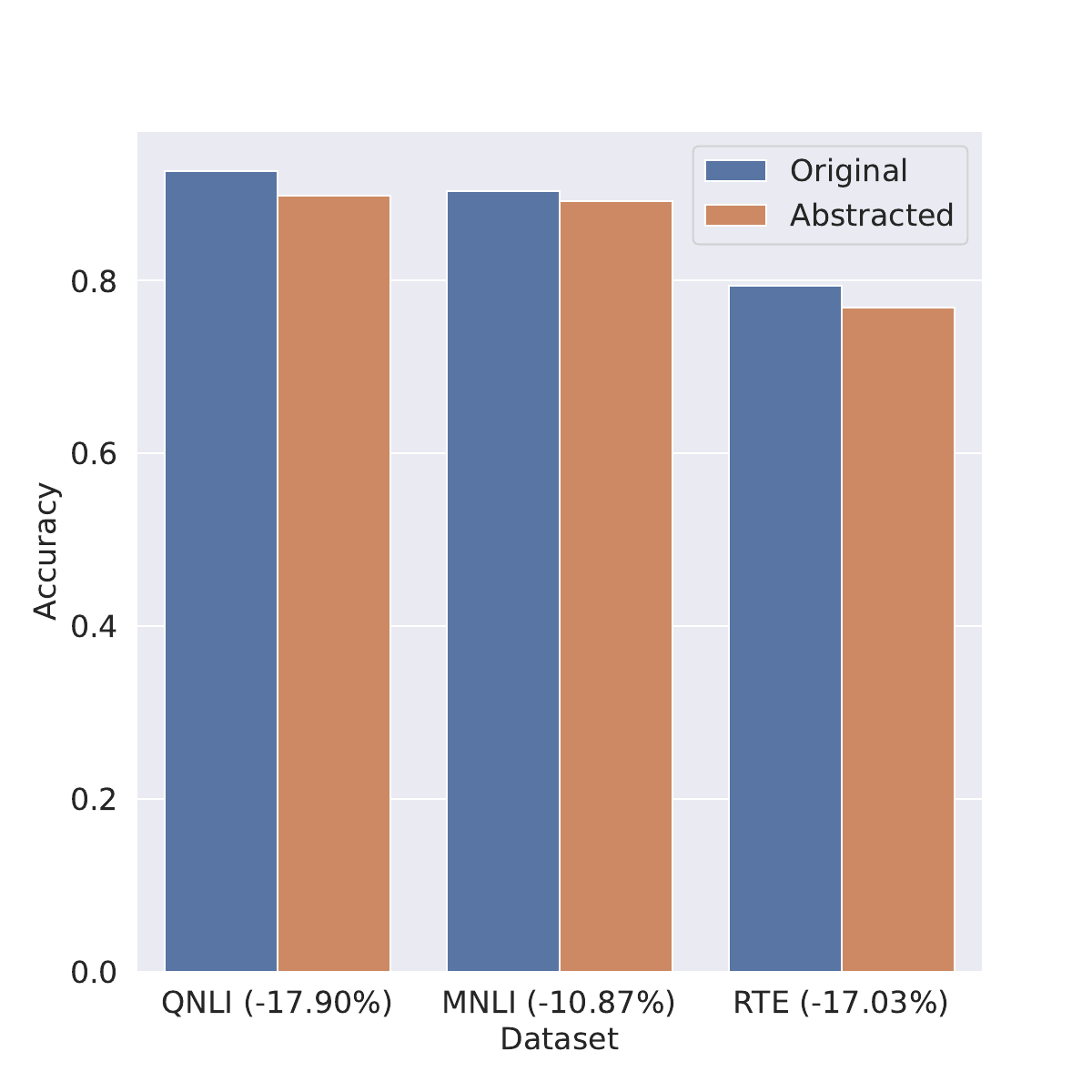}
\caption{Inference Tasks}
\label{fig:nli_exp}
\end{subfigure}
\begin{subfigure}[!t]{0.6\columnwidth}
\centering
\includegraphics[width=\columnwidth]{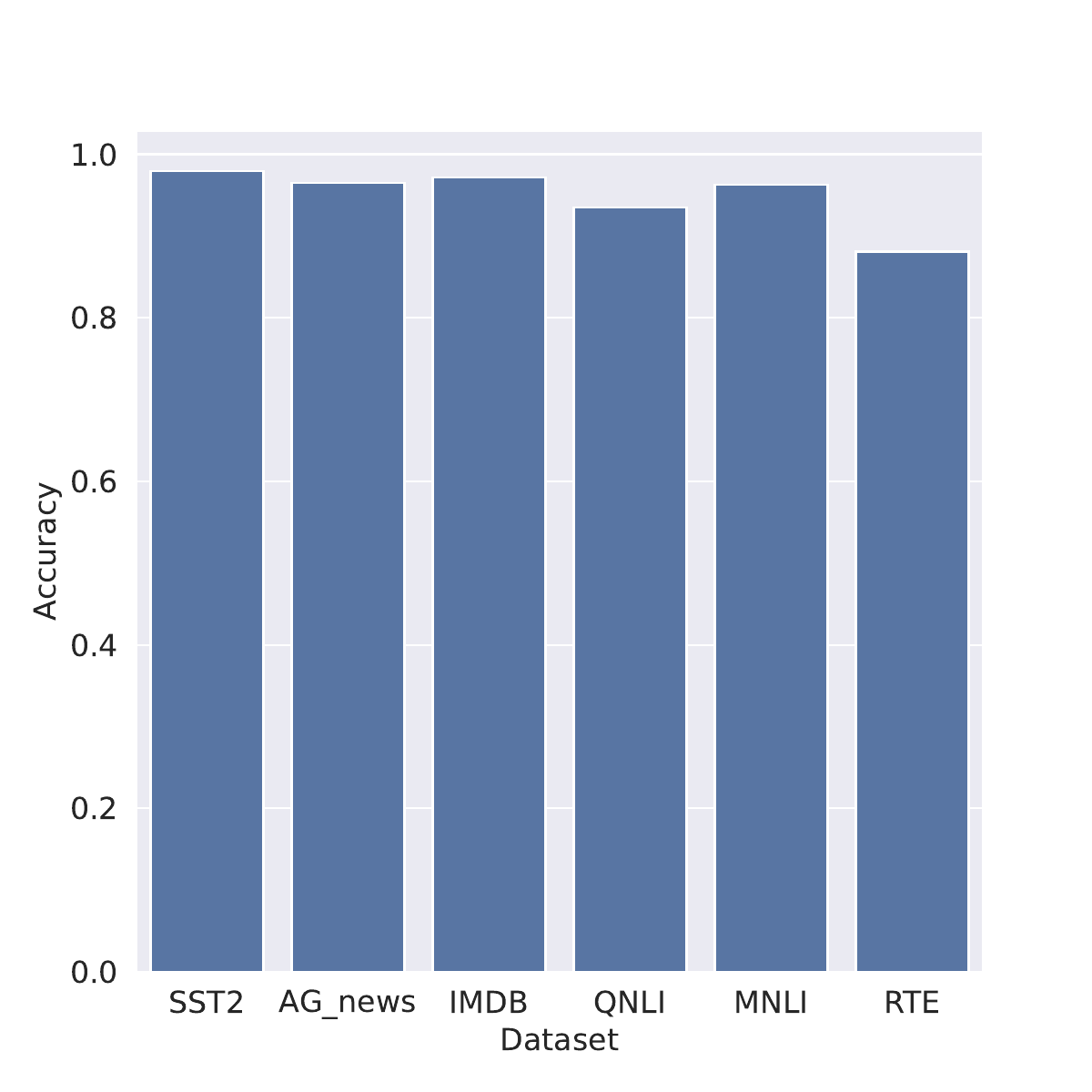}
\caption{Agreement}
\label{fig:cls_agree_exp}
\end{subfigure}
\caption{Utility (Accuracy) and Agreement (Accuracy) of using the original and abstracted classification datasets, and the percentage reduction of sentence (token) length on task-specific models.}
\label{fig:tsne}
\end{figure*}

\subsection{Pesudo API}

In developing the pseudo API, \system requires a tokenizer and a similarity measure function. 
For this, we employ a python package ``tiktoken,''\footnote{\url{https://github.com/openai/tiktoken}} which is developed and released by OpenAI. 
It includes the same tokenizer used in ChatGPT. 
For the similarity measure function, we utilize a pre-trained sentence transformer named ``all-mpnet-base-v2.''~\cite{RG19}
It is important to note that ``all-mpnet-base-v2'' only requires 420 MB of GPU memory, making it significantly smaller in size compared to ChatGPT. 
All experiments are carried out using the PyTorch framework~\cite{PyTorch}.

\subsection{Evaluation Metrics}

To evaluate the performance of \system, we use two metrics: Utility and Agreement.

\mypara{Utility}
Utility measures the performance difference between using original and abstracted inputs on the target model. 
The closer the performance, the better \system.
As we conduct the attack on various NLP tasks, we employ multiple metrics to assess utility.
For the classification task, we evaluate utility based on accuracy. 
In the case of summarization, we compute the F-measure by assessing the degree of overlap between the prediction and reference in terms of unigrams (ROUGE-1), bigrams (ROUGE-2), and the longest sequence of matching terms (ROUGE-L)~\cite{L04}.
Regarding the QA task, we examine the F1 score between the prediction and reference. 
Evaluating the performance of the task-specific model is relatively straightforward, as the output format aligns with the ground truth labels provided by the dataset.
However, it requires more effort for the general language model, and we need to construct templates to generate predictions (refer to the~\autoref{app:templates} for more details). 
In the context of the summarization task, we can evaluate the general language model output directly by employing the ROUGE metric because both the task-specific and general language models are text generation models.
However, the output generated by the general language model for the classification and QA task does not align with the dataset.
In such cases, we determine the label or answer by checking whether it appears within the generated sentence.

\begin{table}[!t]
\centering
\begin{tabular}{lrr}
\toprule
\bf Length & \bf Utility & \bf Agreement \\
\midrule
0\% & 45.05/21.80/30.90 & - \\
-15.21\% & 42.73/17.98/28.53 & 53.11 \\
\bottomrule
\end{tabular}
\caption{Utility (ROUGE-1/2/L) and Agreement (ROUGE-L) of using the original and abstracted CNNDM datasets, and the percentage reduction of sentence (token) length on task-specific models.}
\label{table:cnndm_exp}
\end{table}

\begin{table}[!t]
\centering
\begin{tabular}{lrr}
\toprule
\bf Length & \bf Utility & \bf Agreement \\
\midrule
0\% & 45.51/22.46/37.49 & - \\
-15.26\% & 43.90/20.99/36.05 & 64.58 \\
\bottomrule
\end{tabular}
\caption{Utility (ROUGE-1/2/L) and Agreement (ROUGE-L) of using the original and abstracted XSum datasets, and the percentage reduction of sentence (token) length on task-specific models.}
\label{table:xsum_exp}
\end{table}

\mypara{Agreement}
In addition to assessing utility based on the original dataset, we also evaluate the agreement between responses generated by conditioning on original and abstracted queries. 
For the task-specific model, it is straightforward to use the output based on the original input as a reference and evaluate the agreement using the same metric applied in the utility evaluation.
However, evaluating agreement is crucial for several reasons for the general language model. 
Firstly, since the model is not specifically trained for particular tasks, it might perform worse than the task-specific model. 
Secondly, different templates can impact performance, and inspecting the agreement will mitigate the impact of varying templates. 
Thirdly, the model output is not restricted to a specific format or domain, making it challenging to be evaluated automatically.
For example, the task-specific classification model always outputs a single label that aligns with the ground truth.
In contrast, the general language model can generate the answer within a sentence, making the accuracy calculation more complex.
Besides, instruction prompts usually do not have absolute ground truth as the answer can be open-end.
Therefore, it is more important to check if the target API can generate an appropriate output given the abstracted input.
Instead of solely measuring utility, we quantify the agreement between the general language model's responses using the ROUGE-L metric following~\cite{WBZGYLDDL22, WMAKMNADASPKLPMAKDPPMPPVKVPKDSMAPDS22, MLG23}.
For text generation and instruction tasks, we sampled two outputs with the original input as an upper bound on performance.

\section{Results}
\label{sec:exp_result}

\subsection{Task-Specific Model}
\label{sec:task_exp}

\begin{table}[!t]
\centering
\begin{tabular}{lrr}
\toprule
\bf Length & \bf Utility & \bf Agreement \\
\midrule
0\% & 82.92 & - \\
-15.19\% & 75.19 & 84.10 \\
\bottomrule
\end{tabular}
\caption{Utility (F1) and Agreement (F1) of using the original and abstracted SQuAD2 datasets, and the percentage reduction of sentence (token) length on task-specific models.}
\label{table:sq_exp}
\end{table}

\begin{figure*}[t]
\centering
\begin{subfigure}[t]{0.6\columnwidth}
\centering
\includegraphics[width=\columnwidth]{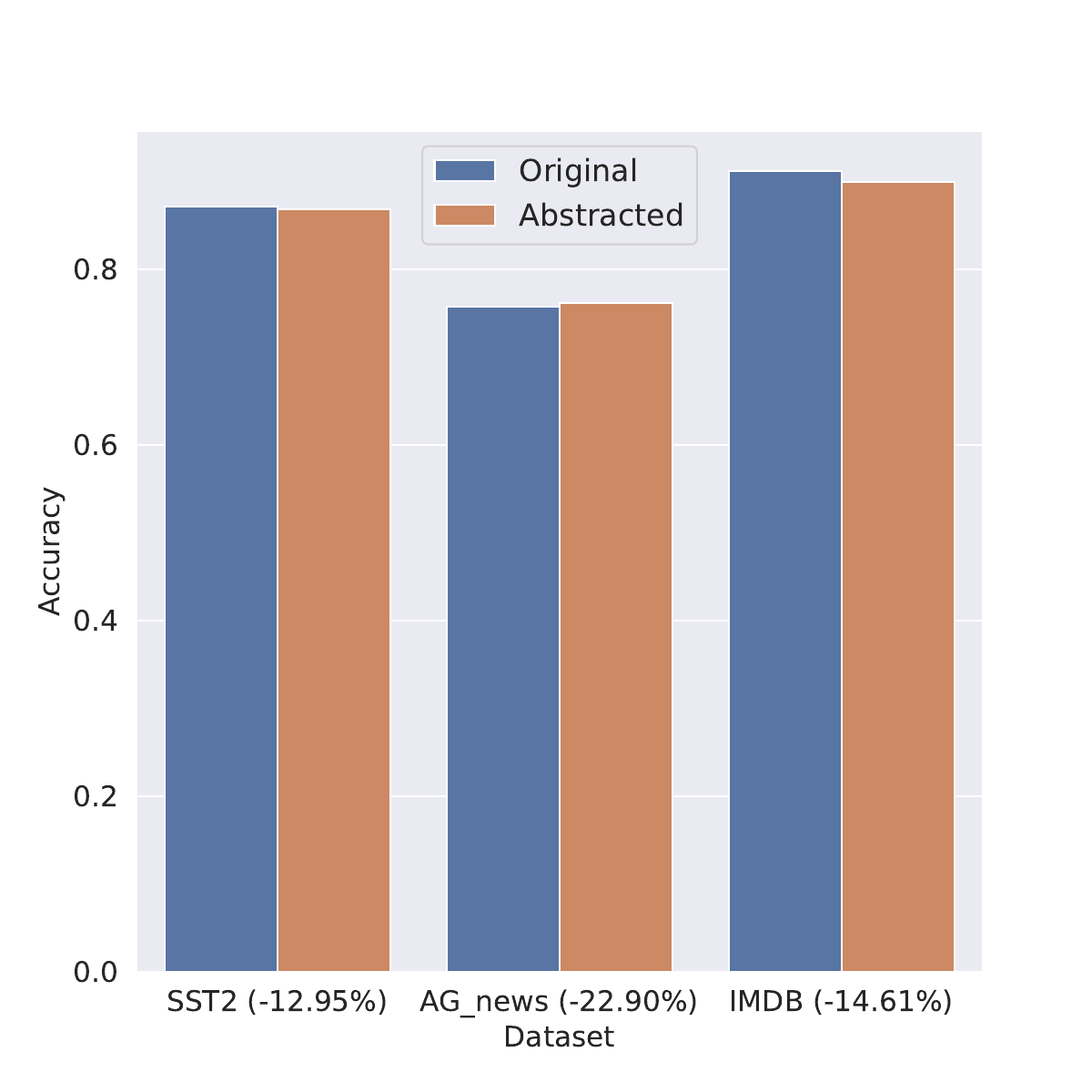}
\caption{Single-Sentence Tasks}
\label{fig:cls_chatgpt_exp}
\end{subfigure}
\begin{subfigure}[t]{0.6\columnwidth}
\centering
\includegraphics[width=\columnwidth]{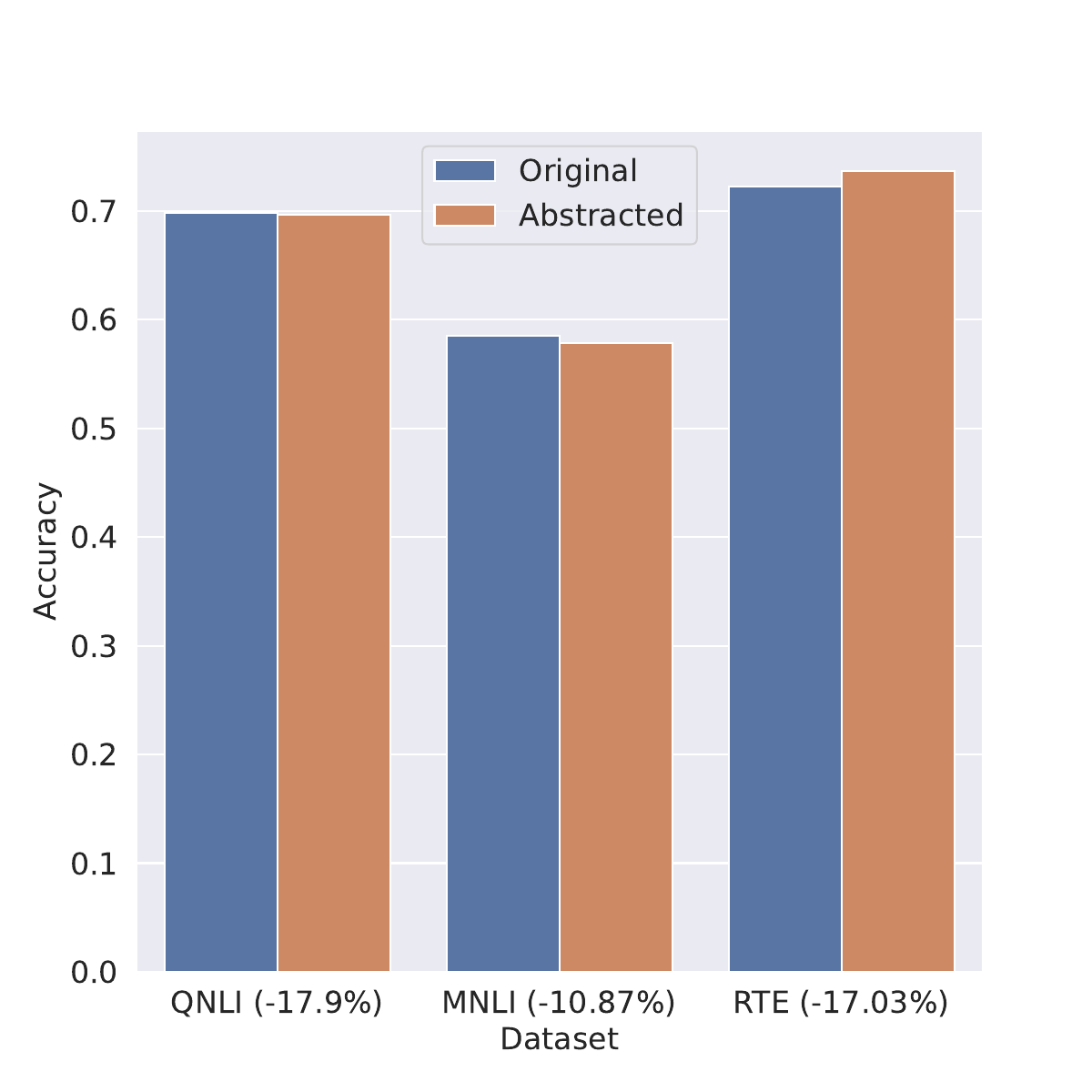}
\caption{Inference Tasks}
\label{fig:nli_chatgpt_exp}
\end{subfigure}
\begin{subfigure}[t]{0.6\columnwidth}
\centering
\includegraphics[width=\columnwidth]{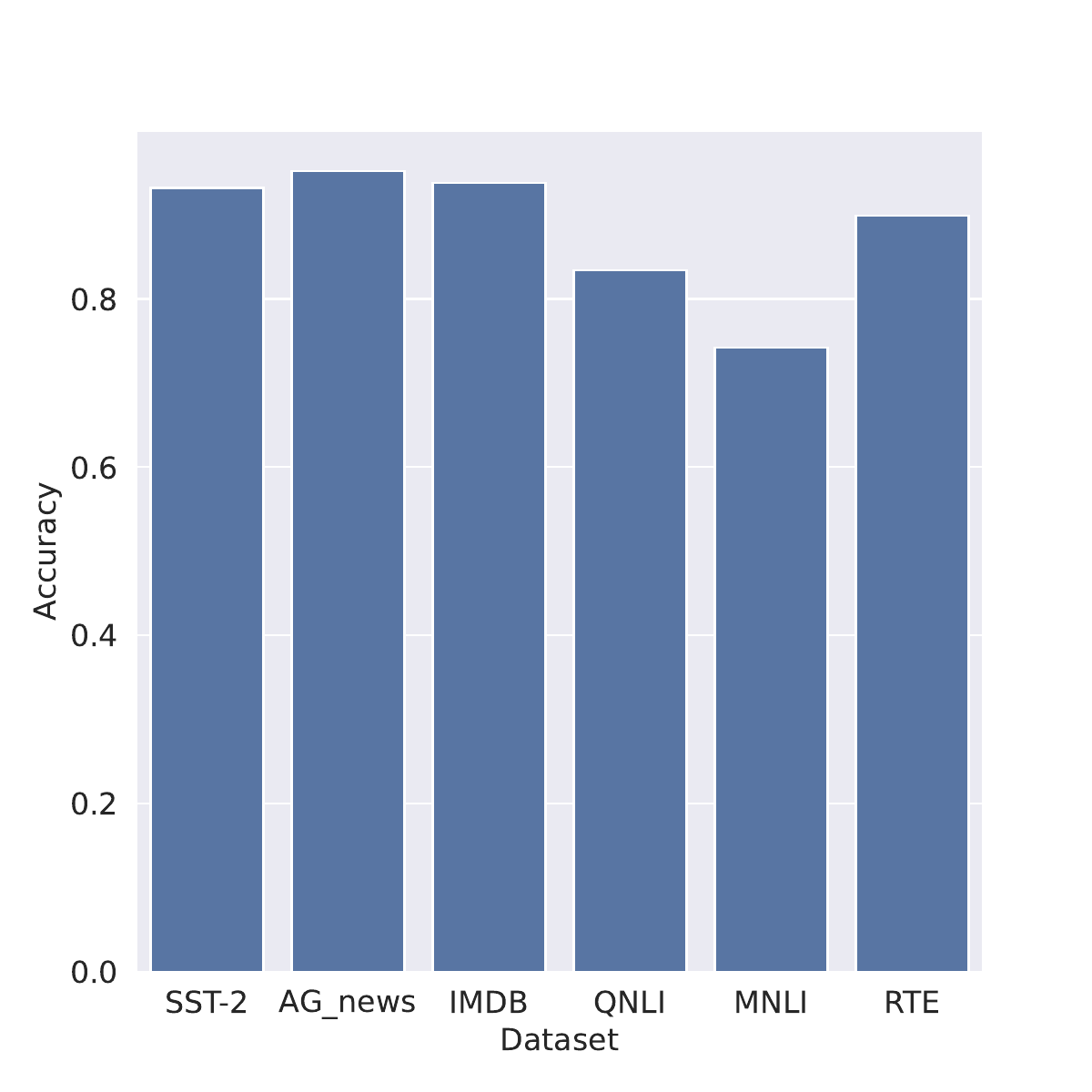}
\caption{Agreement}
\label{fig:cls_chatgpt_agree_exp}
\end{subfigure}
\caption{Utility (Accuracy) and Agreement (Accuracy) of using the original and abstracted classification datasets, and the percentage reduction of sentence (token) length on ChatGPT.}
\label{fig:cls_chatgpt}
\end{figure*}

\mypara{Text Classification}
We start by evaluating the effectiveness of abstracted inputs modified by \system for the classification task. 
The results for three datasets, namely SST-2, AGnews, and IMDB, are presented in \autoref{fig:cls_exp}.
As depicted in the figure, the decrease in utility appears to be negligible, with the decrement being less than 1\%. 
Meanwhile, \system abstracts SST-2, AGnews, and IMDB datasets by 13\%, 23\%, and 15\% in terms of sentence's token length, respectively.
This empirical evidence indicates that abstracted inputs do not jeopardize the utility of the target model on the single-sentence task while reducing the sentence length.
We then examine the output consistency between using original and abstracted inputs. 
The results are represented in \autoref{fig:cls_agree_exp}, which shows an average agreement of 95\%.

\begin{table}[!t]
\centering
\begin{tabular}{lrr}
\toprule
\bf Length & \bf Utility & \bf Agreement \\
\midrule
0\% & 39.82/14.19/25.21 & 39.52\\
-15.26\% & 40.92/15.05/25.82 & 37.44\\
\bottomrule
\end{tabular}
\caption{Utility (ROUGE-1/2/L) and Agreement (ROUGE-L) of using the original and abstracted CNNDM datasets, and the percentage reduction of sentence (token) length on ChatGPT.}
\label{table:cnndm_chatgpt_exp}
\end{table}

\begin{table}[!t]
\centering
\begin{tabular}{lrr}
\toprule
\bf Length & \bf Utility & \bf Agreement \\
\midrule
0\% & 23.57/7.39/16.79 & 36.18 \\
-15.26\% & 23.00/6.65/16.63 & 35.31 \\
\bottomrule
\end{tabular}
\caption{Utility (ROUGE-1/2/L) and Agreement (ROUGE-L) of using the original and abstracted XSum datasets, and the percentage reduction of sentence (token) length on ChatGPT.}
\label{table:xsum_chatgpt_exp}
\end{table}

Next, our investigation explores how the model performs when presented with two abstracted inputs.
This experiment incorporates QNLI, MNLI, and RTE datasets, as shown in \autoref{fig:nli_exp}. 
The results in the figure indicate a minor reduction in utility while reducing the QNLI, MNLI, and RTE sentence's token length to 18\%, 12\%, and 17\%, respectively.
This decrease in utility remains below 3\% accuracy across datasets, and the average agreement is above 90\%.
The result provides additional evidence to support the reliability of abstracted inputs, suggesting the robustness of \system for the classification task.

\mypara{Text Generation}
Second, we evaluate \system with the text generation task. 
In contrast to classification, generation tasks such as summarization require the model not only to comprehend the input but also to produce a coherent output. 
Removing non-trivial words from the input can lead to a decrease in the quality of the generated summary.
Hence, we perform experiments to evaluate the impact of using abstracted inputs in the summarization task.

We conduct the study on two popular datasets: CNNDM and XSum. 
The results are represented in \autoref{table:cnndm_exp} and \autoref{table:xsum_exp}.
As depicted in the tables, using abstracted input does not cause any significant detriment to the model's utility on CNNDM and XSum datasets.
Along with this, there is a reduction of approximately 15\% in the number of tokens on both CNNDM and XSum datasets, indicating the effectiveness of \system.
For the agreement, it is notable to mention that the best ROUGE-L scores achieved with original inputs for CNNDM and XSum datasets are around 31 and 37, respectively. 
Hence, achieving the ROUGE-L scores of 53 and 65 with abstracted inputs is commendable.

\begin{table}[!t]
\centering
\begin{tabular}{lrr}
\toprule
\bf Length & \bf Utility & \bf Agreement \\
\midrule
0\% & 78.19 & - \\
-15.19\% & 73.02 & 75.59 \\
\bottomrule
\end{tabular}
\caption{Utility (F1) and Agreement (F1) of using the original and abstracted SQuAD2 datasets, and the percentage reduction of sentence (token) length on ChatGPT.}
\label{table:sq_chatgpt_exp}
\end{table}

\begin{table*}[!t]
\centering
\setlength\tabcolsep{2.5pt}
\begin{tabular}{lll}
\toprule
 \bf Type & \bf Input & \bf Output \\
\midrule
Original & \begin{tabular}[c]{@{}l@{}}\textbf{What is the} best way \textbf{to} maximize results \\ when using a GPT model?\end{tabular} & \begin{tabular}[c]{@{}l@{}}As an AI language model, I am not capable of having preferences. \\ However, here are some ways to maximize results ...\end{tabular} \\
Abstract & \begin{tabular}[c]{@{}l@{}}best way maximize results when using a \\ GPT model ?\end{tabular} & \begin{tabular}[c]{@{}l@{}}As an AI language model, I don't have experience training or maximizing \\ GPT models or any other AI models, as I am only a pre-trained model. \\ However, here are some general tips to optimize the performance ...\end{tabular} \\
\midrule
Original & \begin{tabular}[c]{@{}l@{}}\textbf{Analyze} the \textbf{given} sentence and explain \\ \textbf{the} context \textbf{in} which the \textbf{sentence} could \\ be applicable \textbackslash{}"The sky is the limit\textbf{\textbackslash{}"}\end{tabular} & \begin{tabular}[c]{@{}l@{}}The sentence "The sky is the limit" means that there are no boundaries \\ or limitations on what one can achieve, accomplish or aspire to. ...\end{tabular} \\
Abstract & \begin{tabular}[c]{@{}l@{}}analyze the sentence and explain context \\ which the could be applicable \textbackslash{}" The sky \\ is the limit\end{tabular} & \begin{tabular}[c]{@{}l@{}}"\textbackslash{}n\textbackslash{}nThe sentence "The sky is the limit" is a popular idiom that is \\ often used to express boundless possibilities. ...\end{tabular} \\
\bottomrule
\end{tabular}
\caption{Examples of the original and abstracted Alpaca prompts and their corresponding outputs from ChatGPT.
We highlight the deleted or transformed words.}
\label{table:prompt_examples}
\end{table*}

\mypara{Question Answering}
Compared to classification and generation, the QA task presents a unique challenge.
First, \system must avoid removing essential words from the context.
Second, the abstracted sentence (question) should remain understandable to the target model.
These considerations make the QA task more challenging than the classification and generation task.
As previously mentioned, the format of the QA task can be classified into two categories: extractive and abstractive.
For the task-specific model, we implement the extractive setup. 

The results of using the abstracted SQuAD2 dataset are represented in \autoref{table:sq_exp}. 
The results show an 8-point (approximately 9\%) decline in the F1 score when using abstracted inputs containing 15\% fewer tokens.
Compared to the classification and generation task, this decrease in utility is relatively higher.
However, given the complexity of the QA task and the length reduction achieved through \system, this level of performance is still acceptable.
Furthermore, an agreement of 85\% is reached between the outputs conditioning on original and abstracted context-question inputs.
These findings further attest to the efficacy of \system, suggesting that it can preserve essential information within both the context and the question.

\mypara{Takeaways}
Our analysis reveals several critical characteristics across different tasks on task-specific models. 
First, abstracting inputs has a minimal impact on the utility when it comes to the classification task. 
This observation aligns with previous research suggesting that only a few important words in a sentence can significantly influence the prediction.
Second, the text generation model demonstrates the robustness of summarizing the context even if it is abstracted.
Third, the QA task is indeed more challenging as it has more utility drops compared to the classification and generation. 
Nevertheless, the utility of the QA model remains competitive.
In summary, our pseudo API powered by \system can deliver decent performance in terms of utility while concurrently offering services at a minimum average of 15\% reduced price.

\subsection{General Langauge Model}
\label{sec:glm_exp}

In contrast to task-specific models, general language models can generate answers for various tasks without fine-tuning. 
However, these models may exhibit inferior performance due to the lack of fine-tuning for specific tasks. 
Therefore, evaluating the agreement becomes more crucial, and we conduct the study against the real-world application, ChatGPT.

\mypara{Text Classification}
Following the same setup as in~\autoref{sec:task_exp}, we extend the evaluation to ChatGPT.
As~\autoref{fig:cls_chatgpt_exp} shows, utilizing abstracted inputs result in only slight reductions in utility for SST-2 and IMDB, with decreases of 2.2\% and 1.4\%, respectively. 
Interestingly, using abstracted inputs has no impact on the performance of AGnews dataset.
Also, the agreement is above 90\% on average, as seen in~\autoref{fig:cls_chatgpt_agree_exp}, demonstrating the robustness of using abstracted inputs from \system on ChatGPT.

\begin{figure}[!t]
\centering
\includegraphics[width=0.6\columnwidth]{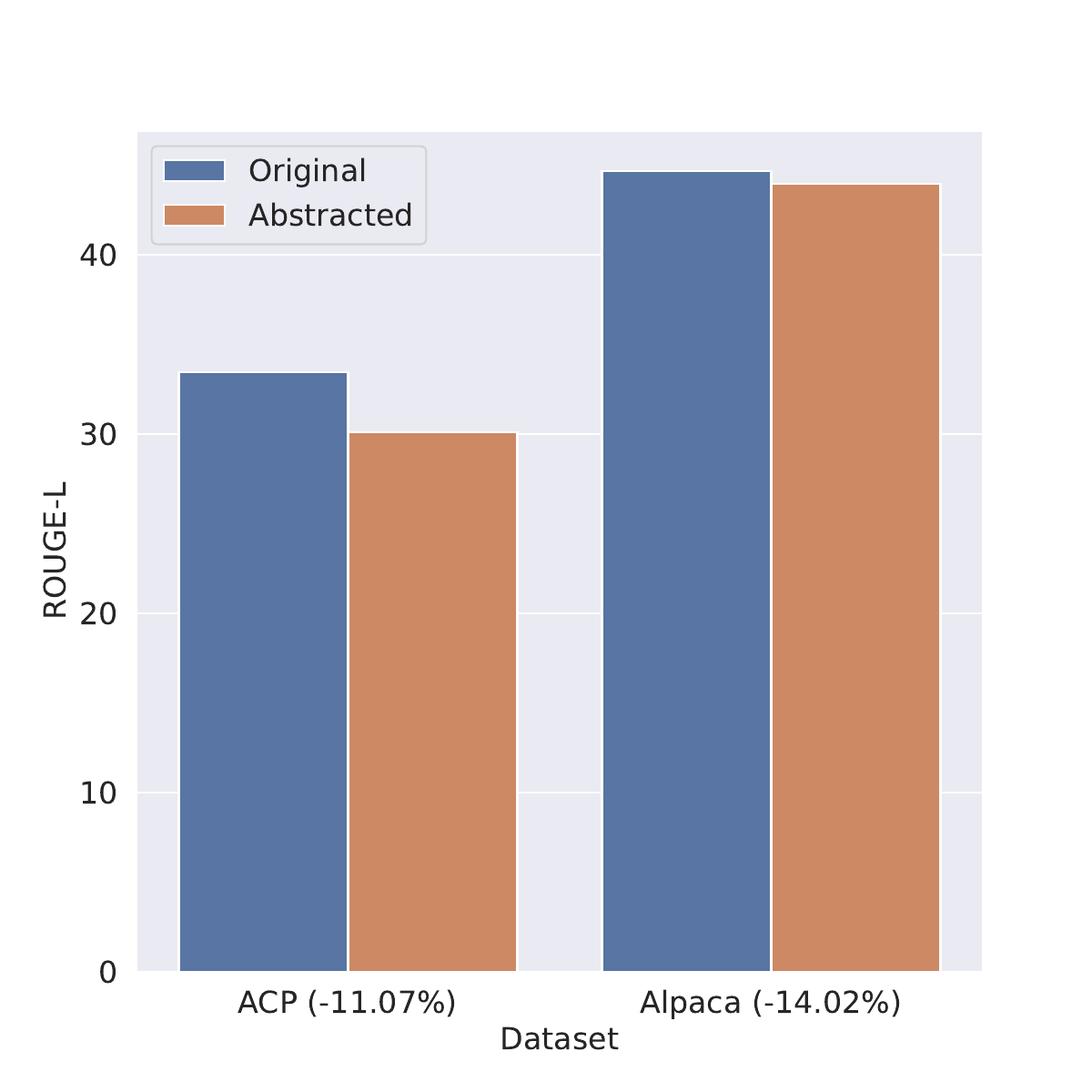}
\caption{Agreement (ROUGE-L) of using the original and abstracted instruction datasets and the percentage reduction of sentence (token) length on ChatGPT.}
\label{fig:prompt_exp}
\end{figure}

For the sentence inference problem, the utility is similar regarding whether original or abstracted inputs are used, as shown in~\autoref{fig:nli_chatgpt_exp}.
Despite the consistency in utility, the agreement in the inference task tends to be relatively lower compared to the single-sentence classification. 
In addition, we observe that the number of labels can influence this drop in agreement.
For instance, datasets with two labels, such as RTE and QNLI, exhibit relatively high agreement levels of 90\% and 83\%, respectively.
Conversely, the MNLI dataset, which incorporates three labels, displays a more substantial drop (74\%) in agreement, highlighting the influence of dataset characteristics on agreement performance.

\mypara{Text Generation}
In terms of summarization performance, the utility also remains consistent whether one uses original input or the abstract version of it, as shown in~\autoref{table:cnndm_chatgpt_exp} and~\autoref{table:xsum_chatgpt_exp}. 
However, it is important to note that evaluating \system using ground truths from the dataset may not accurately reflect its capabilities due to format discrepancies.
For instance, the ground truth of XSum dataset is a single-sentence summary, which contrasts with our prompt template that will produce a short paragraph summary. 
Given this disparity, we emphasize evaluating \system based on the agreement rather than a direct comparison with ground truths.
Compared to the upper bound, we observe less than a 2-point drop in ROUGE-L using abstracted inputs from CNNDM and XSum datasets. 
This minor reduction suggests that our \system is proficient in abstracting user inputs into a format still usable for ChatGPT.

\mypara{Question Answering}
Given that ChatGPT is a generative model, we adopt the generative setup for QA in this section.
However, it is important to note that ChatGPT can generate answers even when they are not explicitly present in the context.
For a fair comparison, we use the prompt template to constrain the model to answer only based on the given context.
As shown in \autoref{table:sq_chatgpt_exp}, querying ChatGPT with abstracted contexts and questions results in a 5-point drop in the F1 score.
This minor decline in performance can be considered insignificant as the substantial reduction in pricing cost achieved through the abstraction process.
Moreover, the agreement evaluation yielded an F1 score of 76, which further attests to the efficacy of our \system on QA data. 

\begin{table}[!t]
\centering
\begin{tabular}{lrr}
\toprule
\bf Operation & \bf Length & \bf Agreement \\
\midrule
Original & 0\% & 44.67 \\
Transform & -4.17\% & 44.73 \\
Delete & -12.89\% & 43.24 \\
Transform + Delete & -14.02\% & 43.94 \\
\bottomrule
\end{tabular}
\caption{Agreement (ROUGE-L) with different operations on Alpaca data.}
\label{table:op_exp}
\end{table}

\begin{table}[!t]
\centering
\begin{tabular}{lrr}
\toprule
\bf Similarty Rate & \bf Length & \bf Agreement \\
\midrule
Original & 0\% & 44.67 \\
0.99 & -14.02\% & 43.94 \\
0.95 & -35.65\% & 39.13 \\
0.90 & -50.23\% & 33.52 \\
\bottomrule
\end{tabular}
\caption{Agreement (ROUGE-L) with different similarity rates on Alpaca data.}
\label{table:sim_exp}
\end{table}

\mypara{Instruction}
While text classification, generation, and QA tasks serve specific purposes, ChatGPT is primarily used by users for general purposes.
To test the performance of our \system on some real-world use cases, we use two relevant datasets, namely ACP and Alpaca.
These datasets include various instruction prompts with inputs that closely mimic real-world usage, which is a valuable and practical study.
Unlike previous tasks, prompts from ACP and Alpaca dataset do not have absolute answers as most are open-end.
Instead, we evaluate output consistency between using original and abstracted inputs.
As depicted in \autoref{fig:prompt_exp}, we observe only minor decreases in ROUGE-L for ACP and Alpaca datasets.
These results are comparable to those achieved using original input, thereby underlining the robustness of our \system and its potential for wide-ranging applications. 

In~\autoref{table:prompt_examples}, we present two samples along with their corresponding outputs from ChatGPT.
From the first sample, it is clear that the word ``What'' is not necessary for questions.
We believe that other ``WH''-words may also not be necessary since a question mark alone can effectively convey the user's intent to the model.
We hypothesize that this phenomenon might extend to other sentence types, such as an exclamation mark alone can replace certain intonation words.
Interestingly, we observe that transforming the first letter of certain words into lowercase can reduce the token length, as demonstrated in the second sample.
We have observed this occurrence with other words in our study as well.
The first example highlights the capacity of large-scale models to comprehend input even when it is abstracted.
Surprisingly, as illustrated by the second example, the vocabulary list of the tokenizer also emerges as a key factor influencing the results.
These findings suggest that there is likely still unexploited attack potential to be discovered.

\mypara{Takeaways}
The comparison between the general language model (ChatGPT) and the task-specific model reveals that the former performs worse than the latter.
Although the general language model has lower utility in classification and QA tasks, it still maintains a decent level of agreement which is the most critical result in this section.
Overall results prove the usability of abstracted inputs on ChatGPT.
In addition, \system proves effective in maintaining consistency and preserving the intended meaning of instruction inputs, with only negligible difference in outputs generated based on original or abstracted inputs.
In general, these findings highlight the usability and effectiveness of \system.

\section{Ablation Study}
\label{sec:ablation}

Now, we investigate various factors that can influence \system performance.
First, we examine the effect of each operation. 
Second, we study the impact of different similarity thresholds. 
Last, we compare the performance when utilizing token length versus character length as the objective. 
In conducting these ablation studies, we use the same Alpaca data from~\autoref{sec:exp_result}, which closely aligns with real-world user usages.

\begin{table}[!t]
\centering
\begin{tabular}{lrrr}
\toprule
\bf Objective & \bf Token Len. & \bf Char Len. & \bf Agreement \\
\midrule
Original & 0\% & 0\% & 44.67 \\
Token & -14.02\% & -4.95\% & 43.94 \\
Character & -9.49\% & -7.90\% & 42.66 \\
\bottomrule
\end{tabular}
\caption{Agreement (ROUGE-L) with different objectives on Alpaca data.}
\label{table:obj_exp}
\end{table}

\mypara{Operation}
In~\autoref{table:op_exp}, we present the impact of employing the Delete and Transform operations independently and their combination.
First, there is no negative effect on performance when utilizing the Transform operation. 
This outcome aligns with expectations since the Transform operation solely involves converting words with more tokens into words with fewer tokens by using its synonyms.
Second, the Delete operation demonstrates a higher degree of efficiency in abstracting the input, albeit with slight reductions in utility. 
The results imply that despite the minor performance trade-off, the Delete operation is more effective for abstracting the input.
Finally, when both operations are employed together, the impact on performance is similar to that observed when utilizing the Delete operation alone while enhancing the length reduction.

\mypara{Similarty Rate}
Next, we evaluate the impact of employing different similarity thresholds, ranging from 0.9 to 0.99.
The adversary can control the level of abstraction in the generated text by adjusting the similarity thresholds. 
In~\autoref{table:sim_exp}, we observed a decrease in utility scores from 44 to 34 in the ROUGE-L score, which indicates that more information was omitted from the sentences and led to reduced comprehension by the model. 
However, it is important to note that a decrease of 11 points coupled with a 50\% reduction in pricing cost is a reasonable trade-off. 
This trade-off can link to OpenAI's approach of offering different models at varying prices based on user demand. 
Similarly, the adversary can offer different pseudo APIs at various pricing options by adjusting the similarity thresholds, thus allowing users to balance cost and utility according to their specific requirements. 
By offering flexibility in pricing options, our pseudo API services can enhance user experience and make it more appealing.

\mypara{Objective}
Besides using the token count for calculating the usage cost, the character count is another pricing unit, and PaLM is based on it.
Therefore, we conduct a study using the character count as the objective.
\autoref{table:obj_exp} shows that fewer tokens in a sentence do not necessarily give us a shorter sentence in terms of character length. 
This discovery emphasizes the importance of selecting different objectives for different target APIs. 
Depending on the API's pricing unit, the choice between token or character count can have varied implications.
Nevertheless, our findings indicate that, whether tokens or characters were used as the objective, \system is able to adapt and consistently demonstrate impressive performance.
The result highlights the flexibility of our \system, proving its effectiveness across different objective types.

\section{Experimental Opertaiton}

\begin{table}[!t]
\centering
\begin{tabular}{lrr}
\toprule
\bf Operation & \bf Length & \bf Agreement \\
\midrule
Original & 0\% & 44.67 \\
Abstract & -14.02\% & 43.94 \\
Abstract + Frag. & -17.01\% & 42.52 \\
\bottomrule
\end{tabular}
\caption{Agreement (ROUGE-L) between using the original, abstracted, and fragmentized Alpaca data.}
\label{table:frag_exp}
\end{table}

In the previous section, we abstract sentences by removing and transforming words, demonstrating the ability of NLP models to comprehend abstracted queries.
Furthermore, recent large language models like ChatGPT have shown robustness, and thus we attempt two experimental operations to reduce sentence length further.

First, we propose the Fragmentize operation, which corrupts words consisting of two or more tokens.
In \autoref{sec:ablation}, we observe that ChatGPT is robust enough to produce decent output even if 50\% of tokens are removed. 
The results show that semantic precision is not always necessary for the model to generate the correct output. 
Additionally, as illustrated in \autoref{sec:methodology}, ``fewer'' is treated as ``few'' and ``er'' instead of a single word. 
We can expect that the model can recover the information and generate the corresponding output even with partial tokens in the input, thereby creating a shorter sentence.

Second, we introduce the Translate operation, which translates certain words from Language A to Language B.
ChatGPT has demonstrated its ability to solve various problems in this paper and other research.
It supports more than 30 languages worldwide and can communicate in different languages.
From an adversary's perspective, we can offer the pseudo API to users in other languages. 
Furthermore, we observe that the token length of some words in Language A can be greater than the same words in Language B.
For example,``\chinese{相似}'' is the Chinese word for ``similar,'' and ``\chinese{相似}'' consists of three tokens while ``similar'' only comprises one token.
As a result, the adversary can offer the pseudo API to other language users, i.e., Chinese, at a lower price by translating certain words in the sentence to achieve a shorter sentence length.

However, incorporating the Fragmentize and Translate operations into our optimization pipeline presents a challenge, as the existing similarity measure model is limited to corrupted and language-mixed sentences.
This limitation can potentially cause optimization failure during the process. 
Therefore, we provide a basic demonstration of these operations and leave the study in the future. 
We describe the experimental setup used in this section in~\autoref{app:operation}.

\begin{table}[!t]
\centering
\begin{tabular}{lrrr}
\toprule
\bf Operation & \bf Length & \bf Agreement & \bf Lang Error \\
\midrule
Original & 100\% & 37.57 & - \\
Abstract & -10.5\% & 35.06 & - \\
Abstract + Tran. & -16.3\% & 32.40 & 5\% \\
\bottomrule
\end{tabular}
\caption{Agreement (ROUGE-L) and language error rate between using the original, abstracted, and translated Alpaca data (Chinese).}
\label{table:translate_exp}
\end{table}

\begin{table*}[!t]
\centering
\setlength\tabcolsep{3pt}
\begin{tabular}{lll}
\toprule
 \bf Type & \bf Input & \bf Output \\
\midrule
Original &  \begin{tabular}[c]{@{}l@{}}\text{Brainstorm} ways reduce energy\\consumption in a home .\end{tabular} & \begin{tabular}[c]{@{}l@{}}Pros of Eating Fast Food:\textbackslash{}n1. Convenience: Fast food is easily \\ available and can be purchased quickly. ...\end{tabular} \\
Fragmentize & \begin{tabular}[c]{@{}l@{}}Brain ways to reduce energy consumption\\in a home.\end{tabular} & \begin{tabular}[c]{@{}l@{}}Pros:\textbackslash{}n\textbackslash{}n1. Convenient and fast: Fast food is prepared quickly \\and can be easily found in most places, ...\end{tabular} \\
\midrule
Original (English) & Name three countries that border Russia. & Norway, China, and Finland. \\
Original (Chinese) & \chinese{说出与}\chinese{俄罗斯}\chinese{接壤的三个国家。} & \chinese{挪威、芬兰、白俄罗斯。} \\
Translate & \chinese{说出与Russia接壤三个国家。} & \chinese{挪威、芬兰、中国。} \\
\bottomrule
\end{tabular}
\caption{Examples of the fragmentized and translated Alpaca prompts and their corresponding outputs from ChatGPT.
The English version is provided for the translated sample as a reference.}
\label{table:more_examples}
\end{table*}

\subsection{Fragmentize}

The purpose of the Fragmentize operation is similar to the Delete operation, but it drops tokens instead of complete words.
In detail, \system employs a masked language model (MLM) to determine whether a particular token in the word can be recovered if omitted. 
\system replaces tokens with a mask token and queries the MLM, then verify if the token appears in the candidate list. 
If so, \system omits the token and repeats the process until no more words can be fragmented. 
The experiment is conducted based on \autoref{fig:prompt_exp}, and the results are presented in \autoref{table:frag_exp}. 
The Fragmentize operation resulted in a 1.4 (ROUGE-L) drop with 2\% fewer tokens.
The results reveal that precise inputs are not necessary for obtaining appropriate outputs.
Furthermore, the adversary can diminish the cost of using the target API, potentially leading to greater profits. 
In addition, we provide a sample in~\autoref{table:more_examples}, where it is apparent that ChatGPT is capable of comprehending a sentence, even when partial tokens are absent.
For example, ``brainstorm'' and ``brain'' convey different meanings, but ChatGPT still manages to generate the correct output.

\subsection{Translate}

The objective of the Translate operation is to translate certain words in a sentence into another language.
To evaluate its performance, we first translate the Alpaca data used in~\autoref{fig:prompt_exp} into Chinese using Google Translate.
Then, we employ a sentence similarity model that supports multiple languages.
However, as multilingual similarity models are not designed for sentences mixed with different languages, we limit our study by translating a maximum of two words per sentence.
At each iteration, we examine whether the cosine similarity of the candidate sentence exceeds the threshold and verify if the candidate sentence retains the same language using a language classifier; otherwise, the target API may respond in an unintended language. 
As shown in~\autoref{table:translate_exp}, the Translate operation results in a 5\% language error, in which ChatGPT responds in English rather than Chinese.
Furthermore, we observe a minor drop in the agreement score of 3 (ROUGE-L) compared to the result of non-translated abstract sentences.
Despite the language errors and minor agreement drops, we further abstract the input by approximately 6\%. 
This highlights the potential applicability of the Translate operation in real-world contexts.
We provide a sample in \autoref{table:more_examples}, demonstrating that ChatGPT can comprehend Chinese sentences mixed with English words and generate the correct response.

\section{Related Works}
\label{sec:related}

\subsection{Large Language Model}

The emergence of Transformer architecture has significantly impacted the field of Natural Language Processing~\cite{VSPUJGKP17, DCLT19, LLGGMLSZ20, BMRSKDNSSAAHKHCRZWWHCSLGCCBMRSA20}.
GPT-2, a widely recognized large language model, has gained substantial popularity during that time~\cite{RWCLAS19}.
As NLP researches continue to evolve, numerous models have been recently introduced.
Models such as Bloom, OPT, and others have gained a lot of attention due to their ability to perform diverse tasks at a high level, even in zero-shot or few-shot settings~\cite{LLGGMLSZ20, WBZGYLDDL22, KGRMI22}.
Consequently, these mature language models have transformed the landscape of the Machine Learning as a Service market.
Popular examples of this transformation include the introduction of OpenAI's ChatGPT and Google's PaLM, which have revolutionized user convenience~\cite{ADFJLPSTBCCCSHMMMORRTXXZAAABBBBBCCCCCCDDDDDDDFFFFGGGa23}.
Meanwhile, various pricing mechanisms are implemented to charge users accordingly, which is exploited by the adversary in this work.

\subsection{Large Language Model Safety}

Unfortunately, the rapid growth of machine learning in the commercial market has also attracted the attention of malicious users~\cite{GAMEHF23, KLSGZH23}.
Traditionally, NLP models suffer from various issues.
For example, they are vulnerable to adversarial attacks, which can degrade model performance with malicious inputs~\cite{BRB18, LJDLW19, JJZS20, ZQZZMHZLS21}. 
Moreover, the adversary can tamper with the training data of target models, causing the model to generate specific outputs when presented with certain triggers.
BadNL~\cite{CSBMSWZ21} is the first backdoor attack against NLP models.
Researchers have studied the robustness of ChatGPT against various modified inputs and different prompts designed to trigger malicious behaviors~\cite{B23, WHHCZWYHYGJZX23, KLSGZH23}. 
ChatGPT also exhibits toxicity in specific conversations, leading to ethical concerns~\cite{DMRKN23}.
Notably, existing attacks primarily target the model itself.
In this work, we propose the first malicious attack against commercial APIs instead and emphasize the need for further study on protecting commercial APIs.

\subsection{Large Model Efficiency}

On the other hand, as the size of language models continues to increase, researchers are actively exploring methods to address this issue without compromising their effectiveness. 
Mu et al.~\cite{MLG23} introduce a technique where the language model is trained to compress prompts into smaller sets of tokens, which can then be reused to enhance computational efficiency. 
Chen et al.~\cite{CZZ23} propose a method for minimizing the cost associated with machine learning APIs by developing an adaptive querying technique. 
This adaptive method selects and utilizes different APIs based on the specific task and dataset. 
With existing methods, model owners can provide high-quality services using fewer computational resources, and users can utilize APIs more cheaply. 
In this paper, we demonstrate that our pseudo API can respond to users' queries while offering the service at an even lower price. 
Furthermore, our pseudo API can provide all the advantages of the target API because the pseudo API acts as a proxy. 
More importantly, the adversary makes a profit without spending any effort on training and maintaining the model.

\subsection{Adversarial Reprogramming}

Ebrahimi et al.~\cite{EGS19} propose a novel attack by reprogramming the ImageNet classification model to execute MNIST and CIFAR-10 tasks.
This attack works by introducing adversarial perturbations to inputs. 
In detail, the adversary can manipulate the model with modified inputs to perform other tasks instead of the original task. 
Hambardzumyan et al.~\cite{HKM21} extend the attack to NLP models. 
They aim to reprogram the masked language model to perform sentiment prediction by incorporating a few trainable embeddings.
However, it requires white-box access to train the embeddings, which is difficult to apply in real-world APIs. 
From an adversarial perspective, black-box reprogramming allows the attacker to build a pseudo API on top of the target API/model. 
Then, the pseudo API can act as an alternative classifier instead of performing the original target's task. 
This potential setup shares the same benefits as our \name attack.

\subsection{Model Hijacking Attack}

The success of a language model requires a huge amount of data, which opens up the possibility of poisoning the model~\cite{BNL12, JOBLNL18, STLLXCS18}. 
Salem et al.~\cite{SBZ22} propose a novel hijacking attack to repurpose the CIFAR-10/CelebA classification model to perform a different task, i.e., MNIST, as defined by the adversary. 
This requires the adversary to build a hijacking dataset by adding adversarial perturbations to the image and poisoning the target model's training set. 
Recently, Si et al.~\cite{SBZS23} extend the attack to the NLP domain and target text generation models. 
Their attack hijacks text generation models instead of classification models and hides the results of the hijacking task in the output sentence. 
Likewise, the adversary can build a pseudo API on top of the poisoned API and offer the functionality of the hijacking task to users. 
This setup also shares the same benefits as our \name attack.

\section{Discussion \& Conclusion}
\label{sec:discussion}

In this paper, we perform the first \name attack against NLP task-specific and general language models.
Our attack targets the pricing mechanism of NLP APIs by abstracting user queries.
Empirical evaluation shows that the adversary can offer the pseudo API with \system more cheaply than the target API while preserving its functionality.
In general, \system achieves a length reduction ranging from 10\% to 23\% across various NLP tasks.
In other words, the adversary can generate profits from it.
Millions of users utilize commercial NLP products and generate countless queries, so even a 1\% profit can be a massive amount.
Moreover, this new type of attack can cause privacy, security, and parasitic computing risks. 
Since the attacker has access to user queries, they may obtain valuable information. 
The attacker can then sell this information for profit or use it maliciously. 
Furthermore, as the attacker has access to both the query and response, they can secretly modify the communications between two parties, such as altering the response into an unsafe response or inserting advertisements within the response. 
Another risk associated with the \name attack is parasitic computing, where the adversary can save on the costs of training and maintaining a large-scale model (target API).

\mypara{Limitation}
Despite the success of the \name attack, we acknowledge some limitations in our work. 
For instance, \system encounters difficulties with specific task's prompts, such as language translation. 
For these tasks, \system should not abstract the sentence that needs to be translated, but it cannot discern which tokens not to be removed. 
Although the Delelet operation, the Transform operation still functions. 
A plausible solution to this problem involves the implementation of a classifier explicitly trained to decide whether a given input should be abstracted or not. 
However, integrating such a classifier requires additional computational resources. 
Nevertheless, the cost of hosting a classifier is significantly smaller than large language models.

\mypara{Future Work}
First, we plan to improve our \system by incorporating linguistic features.
For example, we can employ part-of-speech tagging to prevent the removal of crucial words such as numbers or dates. 
Likewise, we can use entity recognition to identify entities like locations or countries, avoiding accidental removal of this information. 
Second, we aim to expand our work to other applications, such as image generation. 
Alongside the popularity of NLP, computer vision has garnered significant attention, and we anticipate the emergence of various commercial APIs in this field. 
Last, we aspire to broaden the scope of our research to other commercial NLP APIs. 
Specifically, we are interested in studying the behavior of other APIs, such as PaLM, under the \name attack.  
By expanding the range of our research, we hope to gather more insights into potential vulnerabilities in commercial APIs. 
This will enable us to design more robust and resilient systems.

\begin{small}
\bibliographystyle{plain}
\bibliography{normal_generated_py3}
\end{small}

\newpage
\appendix

\section{Experiment Details}
\label{app:appendix}

\subsection{Evaluation Dataset}
\label{app:dataset}

We report the statistics of each evaluation data, including the average sentence length in the word unit and the number of classes in~\autoref{table:data_stat}.

\subsection{Target Model}
\label{app:target}

To mimic the task-specific model scenario where users may upload or fine-tune models.
We use models from Huggingface Hub and report each task's corresponding model~\autoref{table:target_model}.

\subsection{Templates}
\label{app:templates}

For simplicity, we use simple templates that are similar to templates used in in-context learning~\cite{WBZGYLDDL22, KGRMI22}.
Our templates are provided in~\autoref{table:template_examples}.

\subsection{Operation Setup}
\label{app:operation}

For the Fragmentize operation, we use ``distilroberta-base'' as the masked language model and set the top-k to 503.
For the Translate operation, we use ``distiluse-base-multilingual-cased-v1'' as the sentence similarity model.


\begin{table}[!t]
\centering
\small
\tabcolsep 3pt
\begin{tabular}{lrrr}
\toprule
\bf Dataset  & \bf \# Data  & \bf Avg. Len (Total) & \bf \# Class \\
\midrule
SST-2    & 872     & 20.20    & 2        \\
AGnews   & 7,600   & 43.82    & 4         \\
IMDB     & 25,000  & 266.32   & 2         \\
QNLI     & 5,463   & 43.38    & 2         \\
MNLI     & 19,647  & 34.320   & 3         \\
RTE      & 277     & 59.34    & 2         \\
CNNDM    & 13,368  & 785.71   & -         \\
XSum     & 11,332  & 448.75   & -         \\
SQuAD2   & 11,873  & 159.89   & -         \\
ACP      & 137     & 91.44    & -         \\
Alpaca   & 500     & 15.23    & -         \\
\bottomrule
\end{tabular}
\caption{The statistic of the evaluation dataset.}
\label{table:data_stat}
\end{table}

\begin{table}[!t]
\centering
\small
\tabcolsep 3pt
\begin{tabular}{ll}
\toprule
\bf Task  & \bf Model \\
\midrule
SST-2    & gchhablani/bert-base-cased-finetuned-sst2    \\
AGnews   & JiaqiLee/bert-agnews  \\
IMDB     & lvwerra/distilbert-imdb \\
QNLI     & textattack/roberta-base-QNLI \\
MNLI     & roberta-large-mnli \\
RTE      & JeremiahZ/roberta-base-rte \\
CNNDM    & sshleifer/distilbart-cnn-12-6 \\
XSum     & sshleifer/distilbart-xsum-12-6 \\
SQuAD2   & deepset/roberta-base-squad2 \\
\bottomrule
\end{tabular}
\caption{The list of the target model from HuggingFace Hub.}
\label{table:target_model}
\end{table}

\begin{table*}[!t]
\centering
\small
\tabcolsep 3pt
\begin{tabular}{ll}
\toprule
\bf Dataset & \bf Template \\
\midrule
SST-2  & \{Sentence\} \textbackslash n (Positive or Negative) \\
IMDB   & \{Sentence\} \textbackslash n (Positive or Negative) \\
AGnews & \{Sentence\} \textbackslash n (Politics/Sports/Business/Science) \\
\midrule
QNLI & \{Question\} \textbackslash n \{Sentence\} \textbackslash n (Entailment/Not Entailment) \\
MNLI & \{Premise\} \textbackslash n \{Hypothesis\} \textbackslash n (Entailment/Neutral/Contradiction) \\
RTE  & \{Sentence1\} \textbackslash n \{Sentence2\} \textbackslash n (Entailment/Not Entailment) \\
\midrule
CNNDM & \{Article\}\textbackslash n TL;DR: \{Highlights\} \\
XSum  & \{Document\}\textbackslash n TL;DR: \{Summary\} \\
\midrule
SQuAD2 & \{Context\}\textbackslash n Extract answer for: \{Question\} \\
\midrule
ACP & \{Prompt\} \\
Alpaca & \{Prompt\} \{Input\} \\
\bottomrule
\end{tabular}
\caption{A list of templates.}
\label{table:template_examples}
\end{table*}

\end{document}